\documentstyle[12pt]{article}
\def\be{\begin{equation}}
\def\ee{\end{equation}}
\def\bea{\begin{eqnarray}}
\def\eea{\end{eqnarray}}

\def\a{\alpha}
\def\b{\beta}

\begin{document}
\flushright{
\begin{center}
{\bf \Large{Solving Linear Differential Equations: A Novel Approach}}
\end{center}
\begin{center}
{$^a$\it N. Gurappa} \footnote{dr.n.gurappa@gmail.com},
{$^b$\it Abhijit Sen} \footnote{abhijisen@yahoo.co.in},\\
{$^c$\it Rajneesh Atre} \footnote{ darvish.rajneesh@gmail.com} and
{$^b$\it Prasanta K. Panigrahi}\footnote{pprasanta@iiserkol.ac.in}\\
$^a$Priyadarshini College of Engineering, Sullurpet-524121, Andhra Pradesh,\\
$^b$Indian Institute of Science Education and Research-Kolkata, Mohanpur,\\
Nadia-741252, West Bengal, India\\
$^c$Department of Physics, Jaypee Institute of Engineering and Technology,\\
Guna-473226, India
\end{center}
\abstract{We explicate a procedure to solve general linear differential equations,
which connects the desired solutions to monomials $x^{\lambda}$ of
an appropriate degree $\lambda$. In the process the underlying symmetry
of the equations under study, as well as that of the solutions are
made transparent. We demonstrate the efficacy of the method by showing
the common structure of the solution space of a wide variety of differential
equations \textit{viz}. Hermite, Laguerre, Jocobi, Bessel and hypergeometric
etc. We also illustrate the use of the procedure to develop approximate
solutions, as well as in finding solutions of many particle interacting
systems.}

\section{Introduction}

As is evident, the basic building blocks of polynomials and functions
are the monomials. Superposition of finite and infinite number of
these constituents, with appropriate coefficients, respectively leads
to various polynomials and functions. Well known classical orthogonal
polynomials and functions like Bessel, Airy etc., manifest as solutions
of Schr\"odinger equation in quantum mechanics, in the study of waves,
in optics and electromagnetism. 

Apart from finding solutions of various differential equations, considerable
effort has been put into understanding the symmetry properties of
these equations, which throws light on the structure of the solution
space \cite{w}. Here, we illustrate a procedure of solving linear differential
equations, which leads to an exact mapping between the desired polynomials
and functions and the space of monomials \cite{g,guru1}. In this method, the Euler
operator (EO) $D\equiv x\frac{d}{dx}$ plays a crucial role. It makes essential
use of the two properties of the EO. 

1. The monomials are the eigenfunctions of the EO:
\begin{equation}
x\frac{d}{dx}x^{\lambda}\equiv Dx^{\lambda}=\lambda x^{\lambda}.\label{Fundamental}
\end{equation}

2. The Euler operator 'measures' the degree of an operator, irrespective
of its constituents, 
\begin{equation}
[D,O^{d}]=dO^{d}.\label{Degree}
\end{equation}

In the above $O^{d}$ is an operator of `degree' $d$ e.g., $x^{2}$ and
$x^{3}\frac{d}{dx}$ have degree 2, while $\frac{d^{2}}{dx^{2}}$ and
$\frac{d^{2}}{d x^{2}}+\frac{\alpha}{x^{2}}$ have degree -2. 

In the following, we start with the familiar Hermite, Laguerre and confluent
hypergeometric equations and make use of their symmetry properties
to obtain solutions in a form, which exhibits the mapping between the
respective polynomials and the space of monomials. We then explicate
a more general method, applicable to linear differential equations
of arbitrary order and obtain the desired solutions making use
of the above mentioned properties of the Euler operator. The procedure
is explicated through solution of the oscillator problem. Subsequently, earlier solved examples are connected
with the new approach, considering Hermite differential equation
as the example. After illustrating the solution space of the hypergeometric
equation, we tabulate the solutions and the structure of their respective
differential equations for all the functions and polynomials, commonly
encountered in physics literature. The case of the so called quasi
exactly solvable Schr\"odinger equation is then studied, for which
only a part of the spectrum is analytically obtainable. This example
shows the use of the present approach for finding the exact solution,
as well as in the development of approximation schemes. We then proceed
to many-body interacting Calgero-Sutherland systems and show how our
procedure can be profitably explored for finding out the eigenspectra
and eigenfunctions. We then conclude with directions for future work.

\section{Connecting Polynomials with Monomials}

We start with the well studied Hermite differential equation:
\begin{equation}
[-\frac{1}{2}\frac{d^{2}}{dx^{2}}+(D-n)]H_{n}(x)\equiv\hat OH_{n}(x)=0,\label{HermiteDE}
\end{equation}
which has seen the use of traditional series solution method to the
algebraic approach of raising and lowering operators, for finding
the solution \cite{morse}. It is worthwhile to observe that the above equation
has the Euler operator and a degree d=-2 operator $-\frac{1}{2}\frac{d^{2}}{dx^{2}}=P$,
apart from a harmless constant. The fact that $[D,P]=-2P$ can
be profitably used through the Baker-Campbell-Hausdorff (BCH) formula:
\begin{equation}
e^{A}Be^{-A}=B+[A,B]+\frac{1}{2}[A,[A,B]]+...,\label{BCH formula}
\end{equation}

in order to map the differential operator $\hat{O}$ onto (D-n).

For this purpose, it can be straightforwardly checked that
 \begin{equation}
e^{\frac{1}{4}\frac{d^{2}}{dx^{2}}}[-\frac{1}{2}\frac{d^{2}}{dx^{2}}+D-n]e^{-\frac{1}{4}\frac{d^{2}}{dx^{2}}}=-\frac{1}{2}\frac{d^{2}}{dx^{2}}+D-n+\frac{1}{4}[\frac{d^{2}}{dx^{2}},D]=(D-n).\label{Equation01}
\end{equation}

This immidiately suggests to factorize the Hermite polynomial in
the form 
\begin{equation}
[-\frac{1}{2}\frac{d^{2}}{dx^{2}}+D-n]e^{-\frac{1}{4}\frac{d^{2}}{dx^{2}}}\varphi(x)=
e^{-\frac{1}{4}\frac{d^{2}}{dx^{2}}}[D-n]\varphi(x)=0,
\end{equation}
which implies,
\begin{equation}
[D-n]\varphi(x)=0.\label{incomplete2}
\end{equation}
The fact that the monomials are eigenfunctions of the Euler operator
D leads to $\phi(x)=c_{n}x^{n}$ and hence the Hermite polynomial
becomes, $H_{n}(x)=c_{n}e^{-\frac{1}{4}\frac{d^{2}}{dx^{2}}}x^{n}$ \cite{fer}.
The above is the exact connection between the monomial $x^{n}$ and
the polynomial $H_{n}(x)$; the value of $c_{n}$ is to be determined
from normalization condition. The discerning reader can visualize
that the structure of Laguerre differential equation, 
\begin{equation}
[x\frac{d}{dx}-n-(\alpha+1)\frac{d}{dx}-x\frac{d^{2}}{dx^{2}}]L_{n}^{\alpha}(x)=0,\label{Laguerre DE}
\end{equation}

and confluent hypergeometric equation, 
\begin{equation}
[x\frac{d}{dx}+\alpha-x\frac{d^{2}}{dx^{2}}-\gamma\frac{d}{dx}]\Phi(\alpha,\gamma,x)=0,\label{ConflentHypergeometric Diff Eqn}
\end{equation}
are similar to the Hermite case. A similar mapping connects the
above differential operators to $(D-n)$ and hence the solutions can
be written in the form, $L_{n}^{\alpha}=c_{n}exp[-x\frac{d^{2}}{dx^{2}}-(\alpha+1)\frac{d}{dx}]x^{n}$ and
$\Phi(\alpha,\gamma,x)=c_{-\alpha}exp[-x\frac{d^{2}}{dx^{2}}-\gamma\frac{d}{dx}]x^{-\alpha}$.

This procedure can be used to identify
the generating function and the algebraic structure of the solution
space \cite{guru2}. We refer the interested reader to Ref \cite{g} for more details, as also to Ref \cite{TC}
for construction of coherent states for the dynamical systems, associated
with the harmonic oscillator and the Coulomb problem. 

It is evident that, if the differential operator $\hat{O}$ contains
an operator of definite degree, apart from the Euler operator and
the constant term, the aforementioned approach can connect the solution
space to monomials through an exponential mapping. It may not be
the case in general. Keeping this in mind, in the next section we
develop a more general approach connecting the solution space with
monomials for general equations.

\section{A simple approach to familiar differential equations}

For simplicity, we will first consider the case of single variable
linear differential equations (LDEs) and point out its multi-variate generalization later. A
single variable LDE, as will become clear from the examples
of later sections, can be cast in the form
\begin{equation} \label{ie}
\left[F(D) + P(x,d/dx)\right] y(x) = 0, \quad
\end{equation}
where, $F(D) \equiv\sum_{n = - \infty}^{n = \infty} a_n D^n $ and $a_n$'s are some
parameters; $P(x,d/dx)$ can be an arbitrary polynomial function of
$x,d/dx$ and other operators. The solution to Eq. (\ref{ie}) can
be written as \cite{guru1},
\begin{equation} \label{an}
y(x) = C_\lambda \left \{\sum_{m = 0}^{\infty} (-1)^m
\left[\frac{1}{F(D)} P(x,d/dx)\right]^m \right \} x^\lambda,
\end{equation}
provided, $F(D) x^\lambda = 0$; here $C_\lambda$ is constant.
It is easy to see that, the operator $1/F(D)$ is well defined in the
above expression and will not lead to any singularity, if
$P(x,d/dx)$ does not contain any degree zero operator. We note
that $D$ itself is an operator of degree zero.

The proof of Eq.(\ref{an}) is straightforward and follows by
direct substitution \cite{guru1}. Alternatively, since
$F(D)x^{\lambda}=0$, equating $F(D)x^{\lambda}$ modulo
$C_{\lambda}$ and Eq. (\ref{ie}), one finds,
\begin{equation}
\left[ F(D) + P(x,d/dx)\right] y(x) = C_{\lambda} F(D) x^{\lambda} \quad.
\end{equation}
Rearranging the above equation in the form
\begin{equation}
F(D) \left[ 1 + \frac{1}{F(D)} P(x, d/dx) \right] y(x) =
C_{\lambda} F(D) x^{\lambda} \quad,
\end{equation}
and cancelling $F(D)$, we obtain
\begin{equation}
\left[1 + \frac{1}{F(D)} P(x, d/dx) \right]y(x) =
C_{\lambda}x^{\lambda}\quad.
\end{equation}
This yields
\begin{equation}
y(x) = C_{\lambda} \frac{1}{\left[1+
\frac{1}{F(D)}P(x,d/dx)\right]} x^{\lambda}\quad,
\end{equation}
which can be cast in the desired series form: \bea \nonumber y(x)
= C_{\lambda} \sum^{\infty}_{m=0}(-1)^m
\left[\frac{1}{F(D)}P(x,d/dx)\right]^m x^{\lambda}\quad. \eea 
It is explicit that, the above procedure connects the solution $y(x)$
to the space of the monomials $x^\lambda$. 
The generalization of this method to a wide class of many-variable
problems is immediate. Using the fact that, $F(\bar D) X^\lambda =
0$ has solutions, in the space of monomial symmetric functions
\cite{Mac}, where $\bar D = \sum_iD_i \equiv \sum_i x_i
\frac{d}{dx_i}$, the solutions of those multi-variate DEs, which
can be separated into the form given in Eq. (\ref{ie}), can be
solved like the single variable case. As will be seen later, this
procedure enables one to solve a number of correlated many-body
problems.

For illustration, we consider the harmonic oscillator problem. The
Schr$\ddot{o}$dinger eigenvalue equation (in the units,
$\hbar$=$\omega$=$m$=1)
\begin{equation}
\left[\frac{d^2}{d x^2} + (2 E_n - x^2)\right] \psi_n = 0 \quad,
\end{equation}
can be written in the form given in Eq. (\ref{ie}), after
multiplying it by $x^2$ :
\begin{equation}
\left[(D - 1) D + x^2 (2 E_n - x^2)\right] \psi_n = 0 \quad.
\end{equation}
Here, $F(D)=(D - 1)D$ and the condition $F(D) x^{\lambda} = 0$
yields, $\lambda = 0$ or $1$. Using Eq. (\ref{an}), the solution
for $\lambda = 0$ is,
\begin{eqnarray}
\psi_0 &=& C_0 \left \{\sum_{m = 0}^{\infty} (-1)^m
\left[\frac{1}{(D - 1) D} (x^2 (2 E_0 - x^2))\right]^m \right \}
x^0 \nonumber\\ \nonumber\\
&=& C_0 \left[1 - \frac{[2 E_0]}{2 !} x^2 + \frac{(2! + [2
E_0]^2)}{4!} x^4 - \frac{(4! + (2!)^2 [2 E_0] + (2!) [2
E_0]^3)}{2! 6!} x^6 + \cdots \right] \,\,.
\end{eqnarray}
Here $\psi_0$ is an expansion in powers of $x$, whose coefficients
are polynomials in $E_0$. The above series can be written in a
closed, square integrable form, $C_0 \exp(- x^2/2)$, only when
$E_0=1/2$. Analogously, $\lambda=1$, yields the first excited
state. To find the $n^{\mathrm th}$ excited state, one has to
differentiate the Schr${\ddot o}$dinger equation $(n-2)$ number of
times and subsequently multiply it by $x^n$ to produce a $F(D)=
x^n \frac{d^n}{d x^n} = \prod_{l = 0}^{n - 1} (D - l)$, and
proceed in a manner similar to the ground state case.

It is clear that, our procedure yields a series solution, where
additional conditions like square integrability has to be imposed
to obtain physical eigenfunctions and their corresponding
eigenvalues. Once, the ground state has been identified and for
those cases, where $\psi(x)= \psi_0P(x)$, where $P(x)$ is a
polynomial, one can effortlessly obtain the polynomial part, as
will be shown below. Proceeding with the harmonic oscillator case
and writing \bea\nonumber\psi_\a(x) =
\exp(-\frac{x^2}{2})H_{\alpha}(x)\quad,\eea one can easily show
that $H_\a$ satisfies\be \label{h}\left[D - \alpha - \frac{1}{2}
\frac{d^2}{dx^2} \right] H_{\alpha}(x) = 0 \quad, \ee where
$\alpha = E_n-1/2$. The solution of the DE \be \label{shree}
H_{\alpha}(x) = C_{\alpha} \sum_{m = 0}^{\infty}(-1)^m
\left[-\frac{1}{(D - \alpha)}\frac{1}{2} \frac{d^2}{dx^2} \right
]^{m} x^\a \quad.\ee yields a polynomial only when $\a$ is an
integer, since the operator $d^2/dx^2$ reduces the degree of
$x^\a$ by two, in each step. Setting $\a=n$ in Eq. (\ref{h}), we
obtain the Hermite DE and $E_n =(n+1/2)$ as the energy eigenvalue.
Below, we give the algebraic manipulations required to cast the
series solution of Eq. (\ref{shree}) into a form, not very
familiar in the literature. For the Hermite DE, $F(D)=D - n$ and
$P(x,d/dx)=-\frac{1}{2}\frac{d^2}{dx^2}$, the condition $F(D)
x^\lambda = 0$ yields $\lambda = n$, hence,
\begin{eqnarray} \label{sol} H_n(x) = C_n
\sum_{m = 0}^{\infty}(-1)^m \left[-\frac{1}{(D - n)} \frac{1}{2}
\frac{ d^2}{dx^2} \right ]^{m} x^n \quad.
\end{eqnarray}
Using, $[D \,,(d^2/dx^2)] = -2 (d^2/dx^2)$ and making use of the
fact \be \frac{1}{(D-n)} = {\int_0}^\infty ds \, e^{-s(D-n)} \ee
we can write,
\begin{eqnarray} \nonumber
\left[- \frac{1}{2} \frac{1}{(D-n)} \frac{d^2}{dx^2} \right]&=&
- \frac{1}{2} \frac{d^2}{dx^2} \frac{1}{(D-n-2)} \quad, \nonumber \\
\nonumber \\
\left[- \frac{1}{2} \frac{1}{(D-n)} \frac{d^2}{dx^2} \right]
\left[- \frac{1}{2} \frac{1}{(D-n)} \frac{d^2}{dx^2} \right]&=&
\left[ - \frac{1}{2} \frac{d^2}{dx^2}\right]^2 \frac{1}{(D-n-4)}
\frac{1}{(D-n-2)} \quad.
\end{eqnarray}
Hence in general,
\begin{eqnarray} \label{rel}
\left[- \frac{1}{2}\frac{1}{(D - n)}\frac{d^2}{dx^2} \right ]^{m}
x^n &=& \left(- \frac{1}{2}\frac{d^2}{dx^2}\right)^m
\prod_{l=1}^m \frac{1}{(- 2 l)} x^n \quad, \nonumber \\
&=& \frac{1}{m!} \left(\frac{1}{4} \frac{d^2}{dx^2}\right)^m x^n
\quad.
\end{eqnarray}
Substituting Eq. (\ref{rel}) in Eq. (\ref{sol}), we obtain,
\begin{eqnarray}
H_n(x) &=& C_n \sum_{m = 0}^{\infty}(-1)^m \frac{1}{m!}
\left(\frac{1}{4}\frac{d^2}{dx^2}\right)^m
x^n  \quad, \nonumber\\ \label{s}  \nonumber\\
&=& C_n \exp{\left(- \frac{1}{4} \frac{d^2}{dx^2}\right)} x^n \quad,
\end{eqnarray}
a result, not commonly found in the literature. 
The arbitrary constant $C_n$ is chosen to be $2^n$, so that the
polynomials obtained can match with the standard definition
\cite{grad}.

Likewise, the solution to the hypergeometric DE
\begin{equation}
\left[x^2 \frac{d^2}{dx^2} + {\left(\alpha + \beta + 1 \right) x
\frac{d}{dx}} + \alpha\beta - x \frac{d^2}{dx^2} - \gamma
\frac{d}{dx} \right] F{(\alpha, \beta; \gamma; x)}= 0
\end{equation}
can be written as,
\begin{equation}
F(\alpha, \beta; \gamma; x)= (-1)^{-\beta} { \frac{\Gamma ( \alpha
- \beta)\Gamma(\gamma)} { \Gamma (\gamma - \beta) \Gamma(\alpha)}
\exp{\left[\frac{-1}{\left(D+\alpha \right)} \left(x
\frac{d^2}{dx^2}+ \gamma \frac{d}{dx}\right)
\right]}}\,.\,x^{-\beta} \quad.
\end{equation}
For convenience a table has been provided at the end, which lists
a number of commonly encountered DEs and the novel exponential
forms of their solutions. It is worth pointing out that, the
solution of hypergeometric DEs is a polynomial solution, provided either $\a$ or $\b$ are
negative integers. It should be noticed that, unlike the
conventional expressions, the monomials in the above solutions are
arranged in decreasing powers of $x$.

Multiplying the hypergeometric DE with $x$ yields two roots, $\lambda = 0,
1-\gamma$ and $\lambda = 0$ solution gives rise to the well-known,
Gauss hypergeometric series. Since a number of quantum mechanical
problems can be related to confluent and hypergeometric DEs
\cite{landau}, we hope that the novel expressions given above and
 in the table will find physical applications.

\section{Quasi-exactly solvable problems}

This section is devoted to the study of QES problems
\cite{shifman}. These problems are intermediate to exactly and
non-exactly solvable quantum potentials, in the sense that, only a
part of the spectrum can be determined analytically. These
potentials have attracted considerable attention in recent times,
because of their connection to various physical problems \cite{ush}.

We illustrate our procedure, through the sextic oscillator in the
units $\hbar=2m=\omega=1$, for the purpose of comparison with
the standard literature. The corresponding eigenvalue equation is
given by:
\begin{equation}
\left[- \frac{d^2}{dx^2}+ \alpha x^2 +\gamma x^6 \right]\psi(x) =
E \psi(x) \quad.
\end{equation}
 Asymptotic analysis suggests a trial wave function of the form,
\be \psi(x)=\exp(-bx^4){\tilde \psi}(x) \quad,\ee which leads to,
\be\label{sex2}
\left[-\frac{d^2}{dx^2}+2{\sqrt\gamma}x^3\frac{d}{dx}+(\alpha+3{\sqrt\gamma})x^2
 \right]{\tilde \psi}(x)= E\tilde{\psi}(x) \quad,\ee where
 $x^6$ term has been removed by the condition $16b^2=\gamma$.
One notices that, the operator ${\tilde O} =
(\alpha+3{\sqrt\gamma})x^2 +2{\sqrt\gamma}x^3 d/dx$ increases the
degree of ${\tilde \psi}(x)$ by two, if ${\tilde \psi}(x)$ is a
polynomial. Confining ourselves to polynomial solutions and
assuming that the highest power of the monomial in ${\tilde
\psi}(x)$ is $n$, one obtains, \be
-\frac{\alpha}{\sqrt\gamma}=2n+3\ \quad ,\ee  
after imposing the condition that $\tilde{O}$ does not increase the degree of the
polynomial. This is the well-known relationship between the
coupling parameters of the quasi-exactly solvable sextic
oscillator \cite{ush}. Taking $n=4$ and $\gamma=1$ for simplicity, and after
multiplying the above equation with $x^2$ : \be \label{sex2}\left[
D(D-1) + Ex^2 + 8x^4 - 2x^5\frac{d}{dx}\right]{\tilde \psi}(x)= 0
 \quad ,\ee we get,
 \be {\tilde \psi}_{0}(x)=C_0
\left\{ \sum_{m=0}^{\infty}(-1)^m
\left[\frac{1}{D(D-1)}\left(x^2E_0+8x^4-2x^5\frac{d}{dx}\right)\right]^m
\right\}\,1 \quad.\ee Modulo $C_0$, the above series can be
expanded as,\be{\tilde \psi}_{0}(x) = 1 - E_0\frac{x^2}{2!} +
(E_0^2 - 16)\frac{x^4}{4!}+(64E_0 - E_0^3)\frac{x^6}{6!} + \cdots
\quad.\ee The monomials having degree greater than four vanish
provided, $E_0 = 0,\pm 8$. It can be explicitly checked that, for
these values of $E_0$, Eq. (\ref{sex2}) is satisfied. The
eigenfunctions corresponding to these three values are given by,
\bea \psi_{-8}(x) &=&
\exp(-\frac{x^4}{4})[1+ 4x^2 + 2x^4] \\
\psi_0(x) &=& \exp(-\frac{x^4}{4})[1- \frac{2}{3}x^4] \quad,\\
{\mathrm and} \qquad\psi_{+8}(x) &=& \exp(-\frac{x^4}{4})[1- 4x^2
+ 2x^4]\quad.\eea This procedure generalizes to a wide class of
QES problems \cite{atre}.

Below, we demonstrate the method of finding approximate \cite{atre}
eigenvalues and eigenfunctions for non-exactly solvable problems,
using the well studied anharmonic oscillator as the example :
\be\label{anhar1} \left[-\frac{d^2}{dx^2}+\alpha x^2 + \beta x^4 -
E_n\right]\psi_n(x)=0 \quad.\ee Proceeding as before, $\psi_0(x)$
can be written as,
\begin{eqnarray}
\psi_0(x) = C_0 \left \{\sum_{m = 0}^{\infty} (-1)^m
\left[\frac{1}{(D - 1) D}  {\left(E_0 x^2 -\alpha x^4 - \beta
x^6\right)}\right]^m \right \}\,.\, 1 \quad ,
\end{eqnarray} which can be expanded as,
\begin{eqnarray}\label{anhar}
\psi_{0}(x)=1-\frac{
E_0}{2!}x^2+\frac{1}{4!}\left(2\alpha+E_0^2\right)x^4-
\frac{1}{6!}\left(24 \beta-(14\alpha E_0+E_0^3)\right)x^6+
\cdots\end{eqnarray} Although a number of schemes can be devised
for the purpose of approximation, we consider the simplest one of
starting with a trial function ${\tilde\psi}_0(x)= \exp(-\mu
x^2-\nu x^4 )$ and matching it with $\psi_0(x)$. Comparison of the
first three terms yields,
\bea \nonumber \mu&=& \frac{E_0}{2!} \quad,\\
\nonumber \frac{\mu^2}{2!} -
\nu&=&\frac{2\alpha}{4!}+\frac{E_0^2}{4!}\quad,\\
{\mathrm and}\qquad \mu\nu- \frac{\mu^3}{3!}&=& \frac{\beta}{30} -
\frac{(14E_0+E_0^3)}{6!}\quad.\eea The resulting cubic equation in
energy, $E_{0}^{3}- E_0\alpha={3\beta}/2$, leads to one real root
and two complex roots. Choosing the real root on physical grounds,
one obtains,\be E_0=\frac{2^{1/3}\alpha}{A}+\frac{A}{{3}\cdot
2^{1/3}}\quad,\ee where
$A=\left[40.5\beta+(1640.25\beta^2-108\alpha^3)^{1/2}\right]^{1/3}$.
The value of $E_0$, obtained in the weak coupling regime, matches
reasonably well with the earlier obtained results \cite{mfar}. An
approximate $\psi_0$ can be obtained from Eq. (\ref{anhar}). One
can easily improve upon the above scheme by taking better trial
wave functions. Similar analysis can be carried out for the
excited states. The above expansion of the wave function may be
better amenable for a numerical treatment. For example, an
accurate numerically determined energy value can lead to a good
approximate wave function.

\section{Many-body interacting systems}

In this section, we will be dealing with correlated many-body
systems, particularly of the Calogero-Sutherland \cite{calo} and
Sutherland type \cite{suth}. These models have found application
in diverse branches of physics like fluid flow, random matrix
theory, novel statistics, quantum Hall effect and others
\cite{g}. 

We start with the relatively difficult Sutherland model, where the
particles are confined to a circle of circumference $L$. The
two-body problem treated explicitly below, straightforwardly
generalizes to $N$ particles. The Schr\"odinger equation is given
by (in the units $\hbar=m=1$)

\begin{eqnarray} \label{sut1}
\left[- \frac{1}{2}\sum_{i=1}^2 \frac{\partial^2}{ \partial x_i^2}
+ \beta (\beta - 1) \frac{\pi^2}{L^2} \frac{1}{\sin^2[\pi (x_1 -
x_2)/L]} - E_\lambda \right] \psi_\lambda(\{x_i\}) = 0 \qquad.
\end{eqnarray}
Taking, $z_j = e^{2\pi i x_j /L}$ and writing
$\psi_\lambda(\{z_i\}) = \prod_{i,{i \neq j}} z_i^{-\beta
/2}(z_i-z_j)^\beta J_\lambda(\{z_i\})$, the above equation
becomes,
\begin{eqnarray} \label{jac}
\left[\sum_{i=1}^2 D_i^2 + \beta \frac{z_1 + z_2}{z_1 - z_2} (D_1
- D_2) + \tilde{E}_0 - \tilde{E}_\lambda \right]
J_\lambda(\{z_i\}) = 0 \qquad,
\end{eqnarray}
where, $D_i \equiv z_i \frac{\partial}{\partial z_i}$,
$\tilde{E_\lambda} \equiv 2(\frac{L}{2 \pi})^2 E_\lambda$,
$\tilde{E_0} \equiv 2(\frac{L}{2 \pi})^2 E_0$ and $E_0 =
(\frac{\pi}{L})^2 \beta^2$, is the ground-state energy. Here,
$J_\lambda(\{z_i\})$ is the polynomial part, which in the
multivariate case is the well known Jack polynomial \cite{Mac}.
Here, $\lambda$ is the degree of the symmetric function and
$\{\lambda\}$ refers to different partitions of $\lambda$. $\sum_i
D_i^2$ is a diagonal operator in the space spanned by the monomial
symmetric functions, $m_{\{\lambda\}}$, with eigenvalues
$\sum_{i=1}^2 \lambda_i^2$ . A monomial symmetric function is a
symmetrized combination of monomials of definite degree. For
example for two particle case, there are two monomial symmetric
functions having degree two. These are $m_{2,0}=
x_{1}^{2}+x_{2}^{2}$ and $m_{1,1}=x_{1}x_{2}$. Readers are
referred to Ref. [9] for more details about various symmetric
functions and their properties.
 Rewriting Eq. (\ref{jac}) in the form,
\begin{eqnarray}
\left[\sum_i (D_i^2 - \lambda_i^2) + \beta \frac{z_1 + z_2}{z_1 -
z_2} (D_1 - D_2) + \tilde{E}_0 + \sum_i \lambda_i^2 -
\tilde{E}_\lambda \right] J_\lambda(\{z_i\}) = 0 \qquad,
\end{eqnarray}
one can immediately show that,
\begin{eqnarray} \label{sr}
J_\lambda(\{z_i \}) &=& C_\lambda \left\{\sum_{n = 0}^{\infty}
(-1)^n \left[\frac{1}{\sum_i (D_i^2 - \lambda_i^2)}\left(\beta
\frac{z_1 + z_2}{z_1 - z_2}(D_1 - D_2) +  \tilde{E}_0 + \sum_i
\lambda_i^2 - \tilde{E}_\lambda\right)\right]^n \right\} \nonumber\\
&& \qquad \qquad \qquad \qquad \qquad \qquad \qquad \qquad \qquad
\qquad \qquad \qquad  \times  m_\lambda(\{z_i\})\quad .
\end{eqnarray}
For the sake of convenience, we define
\begin{eqnarray} \label{S}
\hat{S} & \equiv & \left[\frac{1}{\sum_i (D_i^2 - \lambda_i^2)}
\hat{Z} \right] \qquad, \nonumber\\
\mbox{and} \qquad \hat{Z} & \equiv & \beta \frac{z_1 + z_2}{z_1 -
z_2} (D_1 - D_2) +  \tilde{E}_0 + \sum_i \lambda_i^2 -
\tilde{E}_\lambda \qquad.
\end{eqnarray}
The action of $\hat{S}$ on $m_\lambda(\{z_i \})$ yields
singularities, unless one chooses the coefficient of $m_\lambda$
in $\hat{Z} \, m_\lambda(\{z_i\})$ to be zero; this condition
yields the eigenvalue equation
$$
\tilde{E}_\lambda = \tilde{E}_0 + \sum_i (\lambda_i^2 + \beta [3-
2 i] \lambda_i) \qquad.
$$
Using the above, one can write down the two particle Jack
polynomial as,
\begin{eqnarray} \label{nfj}
J_\lambda(\{z_i \}) = \sum_{n=0}^\infty (- \beta)^n
\left[\frac{1}{\sum_i (D_i^2 - \lambda_i^2)}(\frac{z_1 + z_2}{z_1
- z_2}(D_1 - D_2) - \sum_i
(3 - 2 i) \lambda_i )\right]^n \nonumber\\
\qquad \qquad \qquad \qquad \qquad  \times m_\lambda(\{x_i\})
\quad.
\end{eqnarray}
Starting from $m_{2,0} = z_1^2 + z_2^2$, it is straightforward to
check that
\begin{eqnarray}
\hat{Z} m_{2,0} &=& 4\beta  (z_1+ z_2)^{2} = 4 \beta m_{1,0}^{2} 
\nonumber\\
\hat{S} m_{2,0} &=& \frac{1}{\sum_i(D_i^2 - 4)}(4 \beta
m_{1,0}^{2})=-2
\beta m_{1,0}^{2} \qquad, \nonumber\\
\hat{S}^n  m_{2,0} &=& - 2 (\beta)^n m_{1,0}^{2} \quad
\mbox{for}\quad n \ge 1 \quad. \nonumber
\end{eqnarray}
Substituting the above result in Eq. (\ref{sr}), apart from $C_2$,
one obtains
\begin{eqnarray}
J_{2} &=& m_{2,0} + \left(\sum_{n=1}^\infty (- 1)^n (-2) (\beta)^n
\right)
m_{1,0}^{2} \nonumber\\
&=& m_{2,0} + 2 \beta \left(\sum_{n = 0}^\infty (- \beta)^n
\right)
m_{1,0}^{2} \nonumber\\
&=& m_{2,0} + \frac{2 \beta}{1 + \beta} m_{1,0}^{2} \qquad;
\end{eqnarray}
which is the desired result. The above approach can be easily
generalized to the $N$-particle case.

Another class of many-body problems, which can be solved by the
present approach is the Calogero-Sutherland model (CSM) and its
generalizations. Proceeding along the line, as for the
Sutherland model, one finds the eigenvalues and eigenfunctions for
the CSM.

The Schr\"odinger equation for the CSM in the previous units, is
given by, \bea
{\left[{-\frac{1}{2}\sum_{i=1}^{N}\frac{\partial}{\partial
{x_{i}}^2} + \frac{1}{2}\sum_{i=1}^{N}x_{i}^{2}+
\frac{g^2}{2}\sum_{{i,j}\atop {i\ne
j}}^{N}{\frac{1}{(x_{i}-x_{j})^2}}}
-E_{n}\right]}\psi_{n}(\{x_i\})=0 \quad,\eea where the wave
function is of the form \cite{calo}, \bea
\psi_{n}(x)=\psi_0P_n(\{x_i\})=ZGP_n(\{x_i\})\quad.\eea Here
$Z\equiv\prod_{i<j}(x_{i}-x_{j})^{\beta}$, $G \equiv
\exp\left\{-\frac{1}{2}\sum_{i}x_{i}^{2}\ \right\}\,$, $g^2 =
\beta(\beta-1)$ and $P_n(\{x_i\})$ is a polynomial. After removing
the ground state, the polynomial $P_n(\{x_i\})$ satisfies,
\begin{eqnarray} \label{til}
\left[\sum_i x_i \frac{\partial} {\partial x_i} +  E_0 - E_n -
\frac{1}{2}\sum_{i}\frac{\partial^2}{\partial{x_{i}}^{2}}-\beta\sum_{i\neq
j}\frac{1}{(x_{i}-x_{j})}\frac{\partial}{\partial{x_{i}}}\right]
P_n(\{x_i\}) = 0 \quad,
\end{eqnarray} where
$E_{0}=\frac{1}{2}N+\frac{1}{2}{\beta}N(N-1)$. Defining \bea
\hat{A}(\beta) \equiv
\frac{1}{2}\sum_{i}\frac{\partial^2}{\partial{x_{i}}^{2}}+\beta\sum_{i\neq
j}\frac{1}{(x_{i}-x_{j})}\frac{\partial}{\partial{x_{i}}} \quad,
\nonumber\eea one can easily see, following the procedure adopted
for the Hermite DE that, \be P_n(\{x_i\}) = C_n
e^{-\hat{A}(\beta)}m_{\{n\}}(\{x_i\}) \quad,\ee where
$m_{\{n\}}(\{x_i\})$ is a monomial symmetric function of degree
$n$. The corresponding energy is given by \bea E_{n}= E_{0}+n
\quad , \eea One can use the above procedure to solve many other
interacting systems.

\section{Conclusions}

In conclusion, we have presented a novel scheme to treat exactly,
quasi-exactly and non-exactly solvable problems, which also
extends to a wide class of many-body interacting systems. The
procedure can be used for the construction of the ladder operators
for various orthogonal polynomials and the quantum systems
associated with them. The approximation scheme presented needs
further refinement. It should be analyzed in conjunction with
computational tools for finding its efficacy as compared to other
methods. The many-body problems presented here have deep
connection with diverse branches of physics and mathematics. The
fact that the procedure employed for solving them, connects the
solution space of the problem under study to the space of
monomials, will make it useful for constructing ladder operators
for the many-variable case. This will throw light on the structure
of the Hilbert space of these correlated systems. Some of these
questions are currently under study and will be reported
elsewhere.

}
\newpage
\noindent{\small {\bf TABLE I. Some frequently encountered DEs and
their novel solutions.}\\
Below, the differential equations from top to bottom,
respectively, are Hermite, Laguerre, Legendre, Gegenbauer,
Chebyshev Type I, Chebyshev Type II, Bessel, confluent
hypergeometric and hypergeometric.
\begin{center}
\begin{tabular}{|c|c|c|}
\hline \hline
{\bf Differential Equation} & {\bf F(D)}, $D\equiv x(d/dx)$ & {\bf 
Solution}\\
\hline &&\\
 $\left[x\frac{d}{dx} - n
-\frac{1}{2}\frac{d^2}{dx^2}\right]H_n(x)=0$ & $(D - n)$ &
$H_n(x)= C_n \exp{\left[-\frac{1}{4}\frac{d^2}{dx^2}
\right]}. x^n$\\
&&\\
 $\left[x\frac{d}{dx}-n -(\alpha+1)\frac{d}{dx}-x\frac{d^2}{dx^2}
    \right]L^{\alpha}_n=0$
    & $(D - n)$
    & $L^{\alpha}_n(x)= C_n\exp{\left[-x\frac{d^2}{dx^2}
    -(\alpha+1)\frac{d}{dx}\right]}. x^n$ \\
&&    \\
$\left[x^2\frac{d^2}{dx^2}+2x\frac{d}{dx}-n(n+1)-\frac{d^2}{dx^2}\right]P_n(x)=0$
&$(D+n+1)(D-n)$ & $P_n(x)= C_n \exp{\left[-\frac{1}{2(D+n+1)}
    \frac{d^2}{dx^2}\right]}. x^n$ \\
$\left[x^2\frac{d^2}{dx^2}+(2\lambda+1) x\frac{d}{dx}-
n(2\lambda+n)-\frac{d^2}{dx^2}\right]$$\times$
&$(D+n+2\lambda)(D-n)$ & $C_n^\lambda(x) = C_n \exp{
    \left[-\frac{1}{2(D + n + 2 \lambda)}
    \frac{d^2}{dx^2}\right]}. x^n$ \\
$\times$$C^{\lambda}_n(x)=0$&& \\
&&\\
$\left[x^2 \frac{d^2}{dx^2}+ x \frac{d}{dx}- n^2- \frac{d^2}{dx^2}
    \right]T_n(x)=0$
&$(D+n)(D-n)$ & $T_n(x)=C_n\exp{\left[-\frac{1}{2(D+n)}
    \frac{d^2}{dx^2}\right]}.x^n$\\
$\left[x^2\frac{d^2}{dx^2}+3x\frac{d}{dx}-n(n+2)-\frac{d^2}{dx^2}\right]U_n(x)=0$
&$(D+n+2)(D-n)$
&$U_n(x)=C_n\exp{\left[-\frac{1}{2(D+n+2)}\frac{d^2}{dx^2}
    \right]}. x^n$\\
&&    \\
$\left[x^2\frac{d^2}{dx^2}+x\frac{d}{dx}- \nu^2 + x^2
\right]J_{\pm \nu}(x)=0$ &$(D+\nu)(D-\nu)$ &$J_{\pm \nu}(x)=C_{\pm
    \nu}\exp{\left[-\frac{1}{2(D \pm \nu)}x^2\right]}. x^{\mp\nu} $\\
&&\\
$\left[x \frac{d}{dx}+\alpha - x \frac{d^2}{dx^2} -
    \gamma\frac{d}{dx}\right] \Phi(\alpha,\gamma,x)=0$
&$(D+\alpha)$ &$\Phi(\alpha,\gamma,x)= C_{-\alpha} \exp{\left[-x
    \frac{d^2}{dx^2}-\gamma \frac{d}{dx} \right]}. x^{- \alpha}$\\
&&    \\
$\left[x(1-x)\frac{d^2}{dx^2}+(\gamma-[\alpha+\beta+1]x)\frac{d}{dx}
    -\alpha\beta\right]\times$
& $(D+\alpha)(D+\beta)$
&$F(\alpha,\beta,\gamma,x)=C_{- (\alpha,\beta)}\times$\\
&&\\
$\times F(\alpha,\beta,\gamma,x)=0$&
&$\times\exp{\left[-\frac{1}{[D+(\alpha,\beta)]}(x\frac{d^2}{dx^2}+\gamma\frac{d}{dx})
    \right]}. x^{-(\beta,\alpha)}$\\
&&\\
\hline \hline
\end{tabular}
\end{center}
The solution to the DE $\left[F(D) + P(x,d/dx)\right] y(x) = 0 $, is,\\
$y(x) = C_\lambda \left\{\sum_{m = 0}^{\infty} (-1)^m
\left[\frac{1}{F(D)}P(x,d/dx)\right]^m \right\} x^\lambda $,
provided $F(D)x^{\lambda}=0$.  \\

\end{document}